\documentclass[aps,twocolumn,pra,reprint,amsmath,amssymb,floatfix,footinbib,superscriptaddress]{revtex4}

\usepackage{amsmath}	%allows the use of mathematical symbols
\usepackage{gensymb}
\usepackage{hyperref}
\hypersetup{colorlinks=true}
\usepackage{upgreek}
\usepackage{graphics}
\usepackage{hyperref}
\usepackage{epsfig}
\usepackage{color}
\usepackage{bm}
\usepackage{float}
\usepackage{ulem} % \sout
\usepackage{blindtext}

\usepackage{CJK}        % For Windows Only
\usepackage{url}
\usepackage{multirow}
\begin{document}
% \begin{CJK*}{GBK}{kai}

\title{High Chern numbers and topological flat bands\\ in high-field polarized Kitaev magnets on the star lattice}

\author{Zixuan Zou}
\affiliation{College of Physics, Nanjing University of Aeronautics and Astronautics, Nanjing, 211106, China}

\author{Qiang Luo\,\href{https://orcid.org/0000-0001-8521-0821}{\includegraphics[scale=0.12]{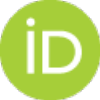}}}
\email[]{qiangluo@nuaa.edu.cn}
\affiliation{College of Physics, Nanjing University of Aeronautics and Astronautics, Nanjing, 211106, China}
\affiliation{Key Laboratory of Aerospace Information Materials and Physics (NUAA), MIIT, Nanjing, 211106, China}

\date{\today}

\begin{abstract}
  The geometrically frustrated Kitaev magnets are demonstrated to be fertile playgrounds that allow for the occurrence of exotic phenomena, including topological phases and the thermal Hall effect. Notwithstanding the established consensus that the field-polarized phase in the honeycomb-lattice Kitaev magnet hosts topological magnons exhibiting Chern numbers $C = \pm1$, the nature of magnon excitations in Kitaev magnets on the star lattice, a triangle-decorated honeycomb lattice, has rarely been explored primarily due to its complicated geometry. To this end, we study the band topology of magnons on the star lattice in the presence of a strong out-of-plane magnetic field using linear spin-wave theory. By calculating the Chern numbers of magnon bands, we find that topological phase diagrams are predominantly composed of two distinct topological phases whose Chern numbers are different by a sign in inverse order. Remarkably, each phase is characterized by a high Chern number of either $+2$ or $-2$. In addition, several topological flat bands with large flatness are identified. The two phases are separated by a dozen narrow topological high-Chern-number segments, whose region shrinks as the magnetic field increases and vanishes eventually. We also find that the thermal Hall conductivity approaches zero at certain parameters, and it changes (keeps) its sign when crossing the topological phase-transition points (flat-band points).
\end{abstract}

%\keywords{}

\maketitle

%%%%%%%%%%%%%%%%%%%%%%%%%%%%%%%%%%%%%%%%%%%%%%%%%%%%%%%%%%%%%%%%%%%%%%%%%%%%%%
\section{Introduction}
In the realm of quantum materials, the search for novel phases exemplified by the Haldane phase \cite{Haldane1983PRL}, Chern insulators \cite{Haldane1988PRL,Hasan2010RMP}, and quantum spin liquids (QSLs) \cite{Anderson1973MRB,Anderson1987Science} has generated an impressive streak in topological characters outside the Landau-Ginzburg-Wilson paradigm.
The Haldane phase is perhaps the earliest example of topological phases which is recognized to have spin-$1/2$ edge modes and nonlocal string order parameter \cite{denNijs1989PRB,Kennedy1992PRB,Pollmann2012PRB}.
Chern insulators are types of topological insulators characterized by nonzero Chern numbers and chiral edge states \cite{Sheng2011NC,Tang2011PRL,Neupert2011PRL}.
In addition, QSLs have highly entangled ground states with topological order, accompanied by fractionalized excitations and emergent gauge structures \cite{Zhou2017RMP,Wen2017RMP,Matsuda2025arXiv}.
Interestingly, it is revealed that vital signatures of these phases can be observed in different Kitaev magnets, ranging from two-dimensional extended Kitaev models \cite{Rau2014PRL,Li2016PRB,Wang2019PRL,Luo2021NPJ,Rousochatzakis2024RoPP,Luo2025CPL} to their spin-chain analogs \cite{Luo2021PRB,Luo2021PRR,Luo2023PRB,Sorensen2024NPJ}.

The Kitaev model, which consists of bond-directional Ising couplings on the honeycomb lattice, was initially proposed in 2006 as a toy model for fault-tolerant quantum computation \cite{Kitaev2006AP}.
It is exactly solvable and hosts QSLs with Majorana fermions and $\mathbb{Z}_2$ gauge fluxes.
Shortly after, with the proposal of the Jackeli-Khaliullin mechanism which provides a route to realize the Kitaev interaction in real compounds with strong spin-orbit coupling \cite{Jackeli2009PRL}, considerable experimental and theoretical advances have been made in synthesizing Kitaev materials and exploring the emergent phenomena in them \cite{Matsuda2025arXiv,Takagi2019NPR,Rousochatzakis2024RoPP}.
Incited by the continuum spectra observed in $\alpha$-RuCl$_3$ by means of inelastic neutron scattering \cite{Banerjee2016NM}, massive experimental work pointed out that there is probably a magnetic-field-induced QSL phase sandwiched between a low-field zigzag order and a polarized phase at a high field \cite{Sears2017PRB,Baek2017PRL,Wolter2017PRB,Wang2017PRL,Zheng2017PRL,Do2017NP,Ran2017PRL,WinterNcom2018}.
This research interest is further fueled by a thermal Hall measurement in which half-quantized thermal Hall conductivity (a signature of Majorana fermions) was reported \cite{Matsuda2018Nature,Matsuda2021Science}.
Although this experiment was not repeated proverbially,
other interesting results, such as oscillations in longitudinal thermal conductivity \cite{Czajka2021NatPhys,Czajka2023NatMater},
are also discerned by experimentalists.
Given such an upsurge of interest, 
the thermal Hall experiments have also been carried out on other Kitaev materials representative of Na$_2$Co$_2$TeO$_6$, Na$_3$Co$_2$SbO$_6$ \cite{Lin2021NC,Li2022PRX,Takeda2022PRR,Guang2023PRB,Gillig2023PRR,Chen2024NC,Fan2025PRB}, and MnPS$_3$ \cite{Yang2024PRB,Nawwar2025arXiv}. However, contributions from phonon excitation and/or spin-lattice coupling are demonstrated to play vital roles in thermal Hall conductivity.

In addition to these widely studied honeycomb compounds, 
materials of geometries with triangle motifs, including triangular, kagome, and star lattices,
have garnered considerable interest because they can enhance quantum fluctuations originating from exchange couplings and geometric frustration of the underlying lattices.
The star lattice, a triangle decorated honeycomb lattice breaking sublattice symmetry, stands out among them as it has the lowest coordination number and largest unit cell, paving the way for the realization of nontrivial emergent phenomena \cite{Yao2007PRL,Dusuel2008PRB}.
Current research endeavors on the star lattice are proceeding along several parallel tracks.
On the one hand, there is keen interest in searching for many-body phases in the Heisenberg model \cite{Yang2010PRB,Ran2018PRB,Jahromi2018PRB,Reingruber2024PRB} and the Kitaev model \cite{dOrnellas2024PRB},
while on the other hand, topological insulators and topological flat bands have been extensively studied in non-interacting spinless fermion systems \cite{Ruegg2010PRB,Chen2012PRB,Chen2012JPCM}.
In addition, routes for the chemical synthesis of star-lattice compounds have been proposed for spin-$1/2$ \cite{Sorolla2020JACS,Ji2022CEJ,Henriques2024NL,Ishikawa2024PRB} or higher \cite{Zheng2007ACIE}.

In the theoretical aspect, the field-polarized Kitaev magnet on the honeycomb lattice is demonstrated to host topological magnons accompanied by Chern numbers $C = \pm 1$ \cite{McClarty2018PRB,Joshi2018PRB}.
Also, topological magnon bands have been realized in the Heisenberg ferromagnet with Dzyaloshinskii–Moriya interaction on the star lattice \cite{Owerre2017JPCM}.
These pose the natural question whether it is attainable to realize topological magnon bands in the field-polarized Kitaev magnet on the star lattice.
To invoke this question further, it is essential to know if there are topological phase transitions, high Chern numbers, and topological flat bands throughout the extensive parameter space.
The aim of this work is to provide affirmative answers to these questions using linear spin-wave theory.

%%%%%%%%%%%%%%%%%%%%%%%%%%%%%%%%%%%%%%%%%%%%%%%%%%%%%%%%%%%%%%%%%%%%%%%%%%%%%%
\section{Model and Method}\label{SEC:Model}
\begin{figure}[!ht]
\centering
\includegraphics[width=0.95\columnwidth, clip]{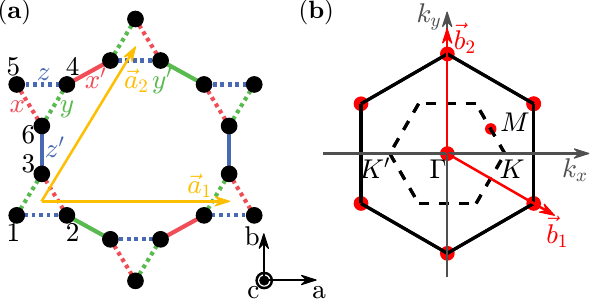}\\
\caption{(a) The structure of the star lattice. Each unit cell contains six spins, labeled as sites 1 through 6.
The dotted (solid) red, green, and blue bonds represent the $x$ ($x'$), $y$ ($y'$), and $z$ ($z'$) bonds, respectively.
We constructed the crystallographic abc axis on the star lattice, with a$[11\bar{2}]$, $b[\bar{1}10]$, and $c[111]$. $\vec{a}_1$ and $\vec{a}_2$ represent the basis vectors within the crystallographic plane.
(b) The first Brillouin zone and the second Brillouin zone of the star lattice. $\vec{b}_1$ and $\vec{b}_2$ are the basis vectors of the reciprocal space.
}\label{FIG-QIMPD}
\end{figure}

We consider the $JK\Gamma\Gamma'$ model with nearest-neighbor interaction on the star lattice in the presence of an external field $\bf{h}$. 
As illustrated in Fig.~\ref{FIG-QIMPD}(a), each unit cell contains six distinct spins labeled from 1 to 6.
There are three inratriangle bonds $\{x, y, z\}$ (dotted lines) and three intertriangle bonds $\{x', y', z'\}$ (solid lines),
which are colored with a label [i.e., red for $x(x')$, green for $y(y')$, and blue for $z(z')$] such that no site connects two bonds of the same color.
For simplicity, we ignore the difference in bond types with or without a prime between intratriangle and intertriangle motifs.
The spin Hamiltonian is given by \cite{Rau2014PRL,Yao2007PRL}
\begin{align}\label{JKGGpHc-Ham}
\mathcal{H} =
    & \sum_{\left<ij\right>\parallel\gamma} \Big[J \mathbf{S}_i \cdot \mathbf{S}_j + K S_i^{\gamma} S_j^{\gamma}
    + \Gamma \big(S_i^{\alpha}S_j^{\beta}+S_i^{\beta}S_j^{\alpha}\big)\Big]   \nonumber \\
    & + \Gamma' \sum_{\left<ij\right>\parallel\gamma}
        \Big[\big(S_i^{\alpha} + S_i^{\beta}\big) S_j^{\gamma} + S_i^{\gamma} \big(S_j^{\alpha} + S_j^{\beta}\big) \Big]    \nonumber \\
    & - \sum_i \mathbf{h} \cdot \mathbf{S}_i,
\end{align}
where $\mathbf{S}_i$ is the spin residing on site $i$, and $S_i^{\gamma}$~($\gamma$ = $x$, $y$, and $z$) is the $\gamma$-component of the $\mathbf{S}_i$ in the cubic $xyz$ axis. 
Meanwhile, $\alpha$ and $\beta$ are determined based on the cyclic permutations of~($\alpha$, $\beta$, $\gamma$)~corresponding to ($x$, $y$, $z$).
In addition, $J$ and $K$ are the diagonal Heisenberg and Kitaev interactions, respectively, while $\Gamma$ and $\Gamma'$ are symmetry-allowed non-diagonal coupling terms.  
$\bf{h}$ represents the magnetic field along the [111] direction, which is strong enough to bring the system into a fully polarized phase.
For convenience, we fix the energy scale $\sqrt{J^2 + K^2 + \Gamma^2 + \Gamma^{'2}} = \mathcal{E}_0$ and adopt a hyperspherical parametrization in which
\begin{equation}\label{EQ:ParaJKGG'}
\left\{
\begin{aligned}
    J       &= \mathcal{E}_0\cos{\theta} \\
    K       &= \mathcal{E}_0\sin{\theta} \cos{\phi} \\
    \Gamma  &= \mathcal{E}_0\sin{\theta} \sin{\phi} \cos{\psi} \\
    \Gamma' &= \mathcal{E}_0\sin{\theta} \sin{\phi} \sin{\psi}
\end{aligned}
\right.
\end{equation}
with $0\leq \theta, \phi, \psi < 2\pi$.

To address quantum fluctuations, we reformulate the Hamiltonian in terms of the Holstein-Primakoff (HP) transformation, utilizing the creation (annihilation) operators $a_i^{\dagger}$ ($a_i$). Within the framework of the linear spin-wave theory, we retain only the quadratic terms (for a review, see Ref. \cite{Janssen2019JPCM}).
The resulting Hamiltonian is then projected into momentum space by employing the Fourier transformation, yielding a quadratic Hamiltonian of the form
\begin{equation}\label{EQ:LSWT}
    \mathcal{H} = C_0 + \frac{S}{2}\sum_{\mathbf{k}}\mathrm{\Psi}^{\dagger}_\mathbf{k}\mathcal{H}_{\mathbf{k}}\mathrm{\Psi}_\mathbf{k}.
\end{equation}
Here, $C_0$ is a non-univerisal value and the Nambu spinor $\mathrm{\Psi}_\mathbf{k}=(a_{1,\mathbf{k}}, a_{2,\mathbf{k}}, \cdots, a_{6,\mathbf{k}}; a^{\dagger}_{1,-\mathbf{k}}, a^{\dagger}_{2,-\mathbf{k}}, \cdots, a^{\dagger}_{6,-\mathbf{k}})^{T}$,
with the operators $a_{n,\mathbf{k}}$ ($n = 1, 2, \cdots, 6$) defined on the six sublattices within a unit cell.
The Bogliubov-de Gennes Hamiltonian $\mathcal{H}_{\mathbf{k}}$ takes the form
\begin{align}
\mathcal{H}_{\mathbf{k}} = 
    \begin{bmatrix}
        \mathcal{A}(\mathbf{k})               &  \mathcal{B}(\mathbf{k})    \\
        \mathcal{B}^{\dagger}(\mathbf{k})     &  \mathcal{A}^T(-\mathbf{k})   \\
	\end{bmatrix}
\end{align}
in which $\mathcal{A}(\mathbf{k})$ and $\mathcal{B}(\mathbf{k})$ are $6 \times 6$ matrices given by
\begin{widetext}
\begin{align}\label{EQ:BdGAB}
\mathcal{A} \!=\! 
    \begin{bmatrix}
        \epsilon_0                  & \lambda_0^1(\mathbf{k})       & \lambda_0^2(\mathbf{k})       & \lambda_0^3(\mathbf{k})       & 0                             & 0 \\
        \lambda_0^{1*}(\mathbf{k})  & \epsilon_0                    & \lambda_0^4(\mathbf{k})       & 0                             & \lambda_0^5(\mathbf{k})       & 0 \\
        \lambda_0^{2*}(\mathbf{k})  & \lambda_0^{4*}(\mathbf{k})    & \epsilon_0                    & 0                             & 0                             & \lambda_0^6(\mathbf{k}) \\
        \lambda_0^{3*}(\mathbf{k})  & 0                             & 0                             & \epsilon_0                    & \lambda_0^{7}(\mathbf{k})    & \lambda_0^8(\mathbf{k}) \\
         0                          & \lambda_0^{5*}(\mathbf{k})    & 0                             & \lambda_0^{7*}(\mathbf{k})    & \epsilon_0                    & \lambda_0^9(\mathbf{k}) \\
         0                          & 0                             & \lambda_0^{6*}(\mathbf{k})    & \lambda_0^{8*}(\mathbf{k})    &\lambda_0^{9*}(\mathbf{k})     & \epsilon_0 \\
	\end{bmatrix},\;
\mathcal{B} \!=\! 
    \begin{bmatrix}
        0                           & \lambda_1^1(\mathbf{k})       & \lambda_1^2(\mathbf{k})       & \lambda_1^3(\mathbf{k})       & 0                             & 0 \\
        \lambda_1^{1}(-\mathbf{k})  & 0                             & \lambda_1^4(\mathbf{k})       & 0                             & \lambda_1^5(\mathbf{k})       & 0 \\
        \lambda_1^{2}(-\mathbf{k})  & \lambda_1^{4}(-\mathbf{k})    & 0                             & 0                             & 0                             & \lambda_1^6(\mathbf{k}) \\
        \lambda_1^{3}(-\mathbf{k})  & 0                             & 0                             & 0                             & \lambda_1^{7}(\mathbf{k})    & \lambda_1^8(\mathbf{k}) \\
         0                          & \lambda_1^{5}(-\mathbf{k})    & 0                             & \lambda_1^{7}(\mathbf{-k})    & 0                    & \lambda_1^9(\mathbf{k}) \\
         0                          & 0                             & \lambda_1^{6}(-\mathbf{k})    & \lambda_1^{8}(-\mathbf{k})    &\lambda_1^{9}(-\mathbf{k})     & 0 \\
	\end{bmatrix}.
\end{align}
\end{widetext}
In Eq.~\eqref{EQ:BdGAB}, the diagonal term is $\epsilon_0 = -[3J+K+2(\Gamma+2\Gamma')] + h/S$, while the off-diagonal terms are 
$\lambda_{\upsilon}^1 = e^{\imath \mathbf{k} \cdot\boldsymbol{\delta}_1} \mathcal{U}_{\upsilon}^z$,
$\lambda_{\upsilon}^2 = e^{\imath \mathbf{k} \cdot\boldsymbol{\delta}_2} \mathcal{U}_{\upsilon}^y$,
$\lambda_{\upsilon}^3 = e^{\imath \mathbf{k} \cdot\boldsymbol{\delta}_3} \mathcal{U}_{\upsilon}^x$,
$\lambda_{\upsilon}^4 = e^{\imath \mathbf{k} \cdot\boldsymbol{\delta}_4} \mathcal{U}_{\upsilon}^x$,
$\lambda_{\upsilon}^5 = e^{\imath \mathbf{k} \cdot\boldsymbol{\delta}_5} \mathcal{U}_{\upsilon}^y$,
$\lambda_{\upsilon}^6 = e^{\imath \mathbf{k} \cdot\boldsymbol{\delta}_6} \mathcal{U}_{\upsilon}^z$,
$\lambda_{\upsilon}^7 = e^{\imath \mathbf{k} \cdot\boldsymbol{\delta}_7} \mathcal{U}_{\upsilon}^z$,
$\lambda_{\upsilon}^8 = e^{\imath \mathbf{k} \cdot\boldsymbol{\delta}_8} \mathcal{U}_{\upsilon}^y$, and
$\lambda_{\upsilon}^9 = e^{\imath \mathbf{k} \cdot\boldsymbol{\delta}_9} \mathcal{U}_{\upsilon}^x$
with $\upsilon$ = 0 [for $\mathcal{A}(\mathbf{k})$] and 1 [for $\mathcal{B}(\mathbf{k})$].
Among these expressions, 
$\boldsymbol{\delta}_1 = (1, 0)$, $\boldsymbol{\delta}_2 = (1/2, \sqrt3/2)$, $\boldsymbol{\delta}_3 = (-\sqrt3/2, -1/2)$,
$\boldsymbol{\delta}_4 = (-1/2, \sqrt3/2)$, $\boldsymbol{\delta}_5 = (\sqrt3/2, -1/2)$, $\boldsymbol{\delta}_6 = (0, 1)$,
$\boldsymbol{\delta}_7 = (-1, 0)$, $\boldsymbol{\delta}_8 = (-1/2, -\sqrt3/2)$, and $\boldsymbol{\delta}_9 = (1/2, -\sqrt3/2)$
are the unit vectors along the bonds in the upper/lower triangles (dotted lines) and hexagons (solid lines) as shown in Fig.~\ref{FIG-QIMPD}.
$\mathcal{U}_{0}^{\gamma}$ and $\mathcal{U}_{1}^{\gamma}$ ($\gamma$ = $x$, $y$, and $z$) are, respectively, given by
$\mathcal{U}_0^{\gamma} = J + \big[K \mathcal{G}_0^{\gamma\gamma} + \Gamma (\mathcal{G}_0^{\alpha\beta}+\mathcal{G}_0^{\beta\alpha}) + \Gamma' (\mathcal{G}_0^{\alpha\gamma}+\mathcal{G}_0^{\gamma\alpha} + \mathcal{G}_0^{\beta\gamma}+\mathcal{G}_0^{\gamma\beta})\big]/2$
and
$\mathcal{U}_1^{\gamma} = \big[K \mathcal{G}_1^{\gamma\gamma} + 2\Gamma \mathcal{G}_1^{\alpha\beta} + 2\Gamma' (\mathcal{G}_1^{\alpha\gamma}+\mathcal{G}_1^{\beta\gamma})\big]/2$, where
\begin{align*}
\mathcal{G}_0 \!=\!
  \frac23\!
  \left[
    \begin{array}{ccc}
     1              & \omega^{-1}       & \omega       \\
     \omega         & 1                 & \omega^{-1}  \\
     \omega^{-1}    & \omega            & 1            \\
    \end{array}
  \right],\;
  \mathcal{G}_1 \!=\!
  \frac23\!
  \left[
    \begin{array}{ccc}
     \omega^{-1}    &   1               & \omega        \\
     1              &   \omega          &  \omega^{-1}  \\
     \omega         &   \omega^{-1}     &  1            \\
    \end{array}
  \right]
\end{align*}
with $\omega = e^{2\pi\imath/3}$.
Here, matrix superscripts follow the Cartesian-index mapping where $(x, y, z) = (1, 2, 3)$, giving the notational equivalence $\mathcal{G}_{\upsilon}^{\alpha\beta} = \mathcal{G}_{\upsilon}^{ij}$ with $\alpha, \beta \in \{x, y, z\}$ and $i, j \in \{1, 2, 3\}$.
Finally, by means of the bosonic Bogoliubov transformation,
the energy-momentum dispersion $\omega_{n,\mathbf{k}}$ of its six magnon bands can be readily determined.
Further, the magnon density of states can be calculated via the retarded Green's function as \cite{Saleem2025PRB}
\begin{align}\label{EQ:DOS}
\textrm{DOS}(\omega_{n,\mathbf{k}}) &= -\frac{1}{\pi} 
\textrm{Im}\int \frac{d^2 \mathbf{k}}{4\pi^2} \textrm{Tr}\left[\frac{1}{\omega_{n,\mathbf{k}} -\mathcal{H}_{\mathbf{k}} +\imath\eta}\right],
\end{align}
where $\eta$ is a positive infinitesimal desrcibing the magnon broadening.
Practically, Eq.~\eqref{EQ:DOS} can be further simplified via the formula $1/(x+\imath\eta) = \textrm{P.V.}\; 1/x -\imath\pi\delta(x)$.

The topological characteristics of magnon bands can be described by finite Chern numbers and large thermal Hall conductivity, which are both expressed as the discrete summation of the weighted Berry curvature over a set of points chosen appropriately spanning the Brillouin zone.
In topologically nontrivial phases, the Berry curvature exhibits significant inhomogeneity in momentum space.
According to the proposal by Fukui \textit{et al.} \cite{Fukui2005JPSJ}, the Chern number associated with the $n$-th band can be efficiently calculated by
\begin{equation}
C_n = \frac{1}{2 \pi} \sum_{\mathbf{k} \in \mathrm{FBZ}} \Omega_{n \mathbf{k}}
\end{equation}
where the Berry curvature is expressed as $\Omega_{n \mathbf{k}} = i F_{12}(k_\ell)$.
In these formulas, $F_{12}(k_\ell)$ is given by $F_{12}(k_\ell) = \mathrm{ln} \left[ U_1(k_\ell) U_2(k_\ell + \hat{1}) U_1(k_\ell + \hat{2})^{-1} U_2(k_\ell)^{-1} \right]$
with $U_\mu(k_\ell) = \langle \phi_n (k_\ell) | \phi_n (k_\ell + \hat{\mu}) \rangle / \mathcal{N}_\mu(k_\ell)$.
Here, $\mathcal{N}_\mu(k_\ell)$ is defined as
$\mathcal{N}_\mu(k_\ell) \equiv \left| \langle \phi_n (k_\ell) | \phi_n (k_\ell + \hat{\mu}) \rangle \right|$,
and $\phi_n$ denotes the eigenvector corresponding to the $n$-th energy band \cite{Fukui2005JPSJ}. 
The discrete lattice points $k_\ell$ ($\ell = 1, \dots, N_1N_2$) are uniformly distributed across the first Brillouin zone (FBZ), 
and $\hat{\mu}$ ($\mu = 1, 2$) represent the unit vectors along the discrete directions within the Brillouin zone [cf. Fig.~\ref{FIG-QIMPD}(b)].
With Chern numbers in hand, the topological phases are characterized by Chern number tuples ($C_1, C_2, C_3, C_4, C_5, C_6)$.
In addition, formula for calculating the thermal Hall conductivity is given by \cite{Matsumoto2011PRL}
\begin{equation}
\kappa_{xy} = -\frac{k_B^2 T}{\hbar V} \sum_{n = 1}^6 \sum_{\mathbf{k} \in \mathrm{FBZ}} c_2 \left[ g(\omega_{n, \mathbf{k}}) \right] \Omega_{n \mathbf{k}}
\end{equation}
where $k_B$ is the Boltzmann constant, $T$ is the temperature, and $V$ is the volume of the system.
$c_2(x) = (1 + x) \mathrm{ln}^2 \left[ (1 + x)/x \right] - \mathrm{ln}^2 x - 2\mathrm{Li} _2 (-x)$, with $\mathrm{Li} _2(x)$ representing the polylogarithmic function and $g(\omega)$ denoting the Bose-Einstein distribution.

\begin{figure*}[htb]
\centering
\includegraphics[width=1.70\columnwidth, clip]{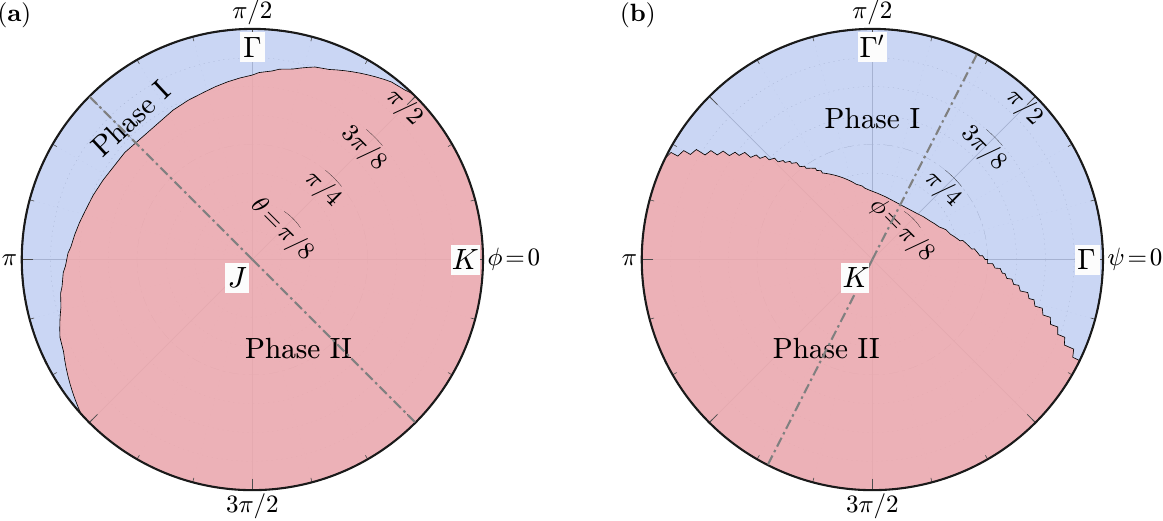}\\
\caption{(a) The topological phase diagram of the $JK\Gamma$ model (with $\psi = 0$) is presented as a function of the parameters $\theta$ and $\phi$. The topological phases are characterized by the Chern numbers of the magnon bands. Specifically, the blue region corresponds to Phase I with a Chern number tuple $(-1, 1, -1, 0, 2, -1)$, while the pink region corresponds to Phase II with a Chern number tuple $(1, -2, 0, 1, -1, 1)$. The dotted gray line marks the symmetric axis $\phi/\pi \simeq 3/4$.
(b) The topological phase diagram of the $K\Gamma \Gamma'$ model (with $\theta = \pi /2$) as a function of $\phi$ and $\psi$. The dotted gray line marks the symmetric axis $\psi/\pi \approx 0.3556$.
  }\label{FIG-PhaseDiag}
\end{figure*}

%%%%%%%%%%%%%%%%%%%%%%%%%%%%%%%%%%%%%%%%%%%%%%%%%%%%%%%%%%%%%%%%%%%%%%%%%%%%%%
\section{Results and Discussion}\label{SEC:IsoKG}
%%%%%%%%%%%%%%%%%%%%%%%%%%%%%%%%%%%%%%%%%%%%%%%%%%%%%%%%%%%%%%%%%%%%%%%%%%%%%%

%%============================================================================
\subsection{Topological phase diagrams at high field}
% %%============================================================================

Within the parameterization framework shown in Eq.~\eqref{EQ:ParaJKGG'}, the relative strengths of these interactions are tuned by exchange coupling angles $\theta$, $\phi$, and $\psi$, facilitating the advent of novel phases and rich phase diagrams.
Here we demonstrate that magnon bands of the fully polarized phase possess nontrivial Chern numbers throughout the entire parameter regions. 
Figures \ref{FIG-PhaseDiag}(a) and \ref{FIG-PhaseDiag}(b) show topological phase diagrams of the $JK\Gamma$ model for tuning $\theta$ and $\phi$, 
and the $K\Gamma\Gamma'$ model for tuning $\phi$ and $\psi$, respectively.
We note that by setting $\theta \rightarrow  \pi-\theta$ and $\phi \rightarrow  \pi+\phi$, there is a global sign change for the interactions $\{J, K, \Gamma\}$ in the $JK\Gamma$ model.
This requires us to consider only $J \geq 0$ here,
and a similar discussion is also applicable to the $K\Gamma\Gamma'$ model.
For comparison, the topological phase diagrams with negative $J$ and $K$ are relegated to Fig. S1 in the Supplemental Material (SM) \cite{SuppMat}. 

At an extremely high magnetic field with $h/(\mathcal{E}_0 S) = 200$,
it is observed from Fig. \ref{FIG-PhaseDiag} that topological phase diagrams of both $JK\Gamma$ and $K\Gamma\Gamma'$ models exhibit only two distinct phases, characterized by the Chern number tuples $(-1, 1, -1, 0, 2, -1)$ (termed phase I) and $(1, -2, 0, 1, -1, 1)$ (termed phase II), suggesting a universal topological feature across these models.
It is also calculated that there are two topological phase-transition points at $\phi/\pi \simeq 1/4$ and $5/4$ ($\psi/\pi \approx 0.8556$ and 1.8556) for the Kitaev-$\Gamma$ model at $\theta = \pi/2$ (for the $\Gamma$-$\Gamma'$ model at $\phi = \pi/2$).
Furthermore, their landscapes are approximately symmetric with respect to lines of $\phi/\pi \simeq 3/4$ and $\psi/\pi \approx 0.3556$, respectively.
In addition, phase II occupies the ground state near the antiferromagnetic Heisenberg limit ($\theta = 0$) or Kitaev ($\phi = 0$) limit. However, there is a difference in the critical exchange coupling angles at which topological phase transitions occur.
The topological phase transitions in the $JK\Gamma$ model are observed exclusively within the interval $\theta/\pi \in [0.3611, 0.5]$, while they are confined to the range $\phi/\pi \in [0.1389, 0.5]$ in the $K\Gamma\Gamma'$ model.

\begin{figure}[!ht]
\centering
\includegraphics[width=0.95\columnwidth, clip]{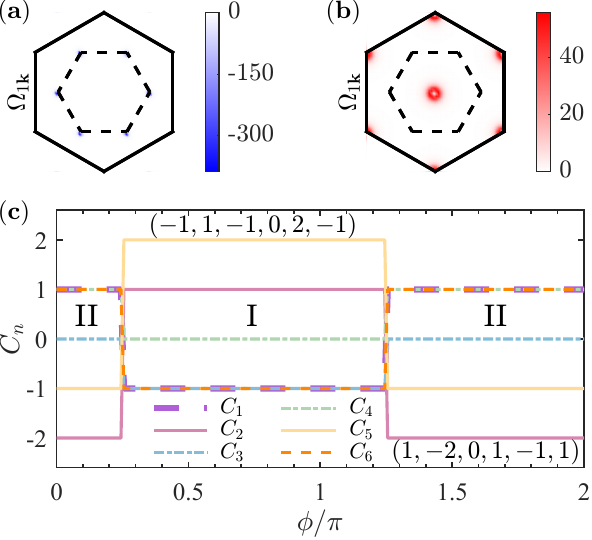}\\
  \caption{Berry curvatures and Chern numbers for the Kitaev-$\Gamma$ model with $h/(\mathcal{E}_0S) = 200$. (a) and (b) are the Berry curvature of the lowest magnon band ($n = 1$) in Phase I for $\phi/\pi = 0.3$ and in Phase II for $\phi/\pi = 0.1$, respectively.
  (c) shows the behaviors of Chern numbers of the six magnon bands.
  }\label{FIG-ChernNo}
\end{figure}

\begin{figure}[!ht]
\centering
\includegraphics[width=0.95\columnwidth, clip]{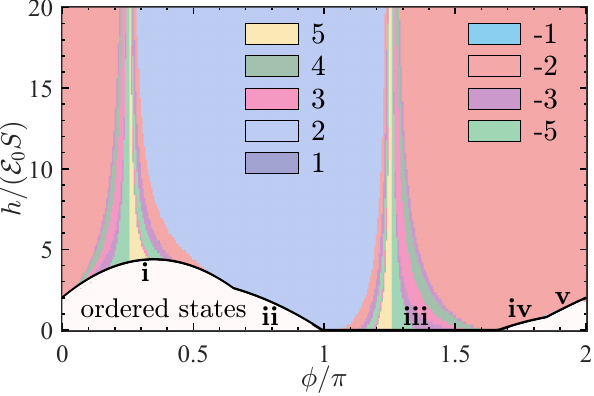}\\
\caption{The topological phase diagram of the Kitaev-$\Gamma$ model in the plane of 
  $[\phi/\pi, h/(\mathcal{E}_0 S)]$ with $h/(\mathcal{E}_0 S) \leq 20$. The white regions represent magnetically ordeed states, while the colored regions denote different topological segments in the field-polarized phase. For each topological phase, it is marked by its higher Chern number regarding absolute value.
  }\label{FIG-PhaseDiagKtvGam}
\end{figure}

%%============================================================================
\subsection{High Chern numbers and topological flat bands}
% %%============================================================================

To delve into the nature of topological phases and their topological phase transitions, in what follows we will focus on the minimal Kitaev-$\Gamma$ model in which $\phi \in [0, 2\pi)$ is the only surviving exchange coupling parameter.
To begin with, we fix the magnetic field $h/(\mathcal{E}_0 S) = 200$ as before and calculate the Berry curvature of the lowest magnon band (i.e., $n = 1$) for $\phi/\pi = 0.3$ and $\phi/\pi = 0.1$, which correspond to phase I and phase II, respectively.
As shown in Figs.~\ref{FIG-ChernNo}(a) and \ref{FIG-ChernNo}(b), 
the Berry curvature peaks respectively near but not exactly at the $\textbf{K}$/$\textbf{K}'$ points and $\boldsymbol{\Gamma}$ point, showing somewhat incommensurate behaviors.
It is negative (positive)  throughout the Brillouin zone for $\phi/\pi = 0.3$ (0.1), giving rise to a Chern number of $-1 (+1)$.
To proceed, Fig.~\ref{FIG-ChernNo}(c) presents the Chern number distribution of the Kitaev-$\Gamma$ model in the range $\phi \in[0, 2\pi)$.
The parameter region is uniformly divided into two parts, in which $(\pi/4, 5\pi/4)$ belongs to phase I while the remaining is for phase II.
The Chern number distribution of the magnon bands in the two phases follows a pattern of inverted order with opposite signs, and the Chern numbers of the lowest and highest bands remain the same.
Notably, both phases exhibit high Chern numbers,
in which the fifth band of phase I exhibits a Chern number of $C_5 = 2$ while the second band of phase II carries $C_2 = -2$.

Next, as we decrease the magnetic field to a modest level, the quantum fluctuations are enhanced, leaving the possibility of generating other topological phases located in the narrow slot along the phase boundaries of phases I and II.
We degrade the discussion of the topological phase transitions with $h/(\mathcal{E}_0 S)$ = 150, 100, and 50 to Fig. S2 in the SM \cite{SuppMat},
and we show the topological phase diagram of the Kitaev-$\Gamma$ model in the plane of $[\phi/\pi, h/(\mathcal{E}_0 S)]$ with $h/(\mathcal{E}_0 S) \leq 20$ in Fig.~\ref{FIG-PhaseDiagKtvGam}.
The regions of ordered states (in white) in the low-field regime are determined from the gap closure in the spin-wave spectrum.
Due to the large unit cell, an exact solution of the magnetic phase boundary seems unachievable.
Nevertheless, we find that it can be approximately fitted by five piecewise quadratic functions of the form
% \begin{equation}\label{EQ:PhaseBound}
% h/S = \left\{
% \begin{array}{lcl}
% -19.54(\frac{\phi}{\pi})^2 + 13.64\frac{\phi}{\pi} + 2,&      & {0 < \phi/\pi \leq 0.6511}\\
% -8.77(\frac{\phi}{\pi})^2 + 6.77\frac{\phi}{\pi} + 1.91,&      & {0.6511 < \phi/\pi \leq 0.9917}\\
% 0,&      & {0.9917 < \phi/\pi \leq 1.6661}\\
% -7.51(\frac{\phi}{\pi})^2 + 30.77\frac{\phi}{\pi} -30.41,&      & {1.6661 < \phi/\pi \leq 1.8497} \\
% -4.46(\frac{\phi}{\pi})^2 + 25.17\frac{\phi}{\pi} - 30.49,&      & {1.8497 < \phi/\pi \leq 2}
% \end{array} \right..
% \end{equation}
\begin{equation}\label{EQ:PhaseBound}
h/S = \lambda \left(\frac{\phi}{\pi}\right)^2 + \mu \frac{\phi}{\pi} + \nu,
\end{equation}
in which the coefficients are listed in Table~\ref{Tab-FitingFormula}.

\begin{table}[th!]
\caption{\label{Tab-FitingFormula}
Fitting coefficients for the quadratic function shown in Eq.~\eqref{EQ:PhaseBound} at five different ranges. The corresponding segments are marked in Fig.~\ref{FIG-PhaseDiagKtvGam}.}
\begin{ruledtabular}
\begin{tabular}{ c c c  c  c}
No.     & $\lambda$     & $\mu$     & $\nu$                 & Parameter Ranges \\
\colrule
i       & $-19.54$      & 13.64     & $\phantom{-}$2.00     &   $\phi/\pi \in (0, 0.6511)$          \\
ii      & $-8.77$       & 6.77      & $\phantom{-}$1.91     &   $\phi/\pi \in (0.6511, 0.9917)$     \\
iii     & 0             & 0         & 0                     &   $\phi/\pi \in (0.9917, 1.6661)$     \\
iv      & $-7.51$       & 30.77     & $-30.41$              &   $\phi/\pi \in (1.6661, 1.8497)$     \\
v       & $-4.46$       & 25.17     & $-30.49$              &   $\phi/\pi \in (1.8497, 2)$          \\
\end{tabular}
\end{ruledtabular}
\end{table}

On top of the magnetic phase boundary, it is surprising to find that there are dozens of intermediate topological phases sandwiched between phase I and phase II.
For better visualization, we denote each phase by its highest Chern number in the sense of absolute value. If the maximum and minimum Chern numbers have identical magnitudes but opposite signs, we adopt the one associated with the lower band. 
As can be seen in Fig.~\ref{FIG-PhaseDiagKtvGam}, the intermediate topological phases are narrow and elongated, and their ranges gradually shrink as the magnetic field increases.
Depending on the values of the exchange coupling angle and magnetic field, their highest Chern numbers vary from $-5$ to $+5$.
Of note is that the regions with the highest Chern numbers $|C| = 5$ are generated on the brick of $\phi/\pi = 1/4$ and $5/4$.
Recalling that in the field-polarized phase on the honeycomb lattice, there is only one topological phase when the field is normal to the honeycomb plane \cite{McClarty2018PRB,Joshi2018PRB} and topological phase diagrams are independent of the field strength in the in-plane field \cite{Chern2024PRB}, our result thus emphasizes that the star lattice can exhibit successive field-induced topological phase transitions.

\begin{figure}[!ht]
\centering
\includegraphics[width=0.95\columnwidth, clip]{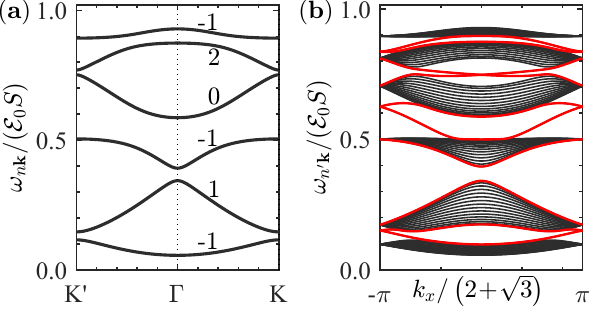}\\
\caption{(a) Magnon band structures for the Kitaev-$\Gamma$ model at the parameter 
  $[\phi/\pi, h/(\mathcal{E}_0 S)]$ = (0.98, 0.30).
  (b) The corresponding zigzag magnon edge modes with sixteen unit cells along the $b$ direction.
  }\label{FIG-EdgeMode}
\end{figure}

Regardless of the specific values of the Chern numbers, their nonzero nature inherently signifies the presence of chiral edge states. Owing to the bulk-edge correspondence, the gapped band structures host in-gap edge states, which are robust against local perturbations due to their topological protection.
Further, pairs of edge states in the $n$th band gap equal the winding number $\mathcal{W}_n$, defined as the sum of Chern numbers up to the current band, i.e., $\mathcal{W}_n = \sum_{n'}C_{n'}$ \cite{Zhuo2022NJP,Li2023CPL,Chen2024PRB,Zhang2023PRB}.
Figure~\ref{FIG-EdgeMode} shows the bulk magnon bands and their corresponding band structures in a strip geometry consisting of sixteen unit cells along the $b$ axis and infinitely long towards the $a$ axis.
Aligning with the fully polarized phase, the lowest magnon gap shown in Fig.~\ref{FIG-EdgeMode}(a) is located at the $\boldsymbol{\Gamma}$ point,
and the energy bands are well separated at the selected parameters $[\phi/\pi, h/(\mathcal{E}_0 S)]$ = (0.98, 0.30) in phase I.
Furthermore, a pair of edge states is clearly observed in all the band gaps except for the second, as shown in Fig.~\ref{FIG-EdgeMode}(b).
This intriguing phenomenon is consistent with the fact that $|\mathcal{W}_n| = 1$ ($n \neq 2$) and $\mathcal{W}_2 = 0$.

Beyond band topology, flat bands, characterized by nearly dispersionless energy-momentum relations that yield vanishing group velocities and high density of states, have attracted considerable research interest in recent years (for reviews, see Refs.~\cite{Liu2014CPB,Rgim2021AIPX}).
Flat-band structures suppress quasiparticle kinetic energy, thereby enhancing the influence of the lattice environment, which synergizes with interactions, perturbations, and particle statistics.
Therefore, the interplay of geometrically frustrated lattice and localization-enhanced correlation effects not only facilitate the emergence of exotic quantum states, but also provide an ideal platform for investigating low-energy physical phenomena.
The extensively studied electronic flat bands, especially those arising from destructive interference in kagome-lattice materials like Fe$_3$Mn$_2$ and CoSn \cite{Lin2018PRL,Liu2020NC}, have enabled explorations of novel phases such as ferromagnetism, superconductivity, and the fractional quantum anomalous Hall effect \cite{Liu2014CPB,Rgim2021AIPX,Hu2023CP}.
However, in bosonic systems involving magnons or others, flat bands can cause fundamentally different novel phenomena because of distinct quantum statistics,
such as supersolid phase \cite{Moller2012PRL}, quantum many-body scar \cite{Kuno2020PRB}, symmetry-protected topological phase \cite{Yang2021PRR}, and Bose-Einstein condensate (BEC) \cite{Wang2024arXiv}.
Notably, in frustrated magnets like SrCu$_2$(BO$_3$)$_2$, flat magnon bands under high magnetic fields lead to a magnon BEC, manifesting as a magnetization plateau.
The high density of states in flat bands elevates the BEC critical temperature, allowing condensation under experimentally accessible conditions \cite{Wang2024arXiv}.
Therefore, motivated by the successful realization of flat bands in the star-lattice tight-binding models \cite{Ruegg2010PRB,Chen2012PRB,Chen2012JPCM} and the Hubbard model \cite{Nourse2022PRB}, we therefore focus on investigating magnon-induced flat-band structures on the star lattice.

\begin{figure}[!ht]
\centering
\includegraphics[width=0.95\columnwidth, clip]{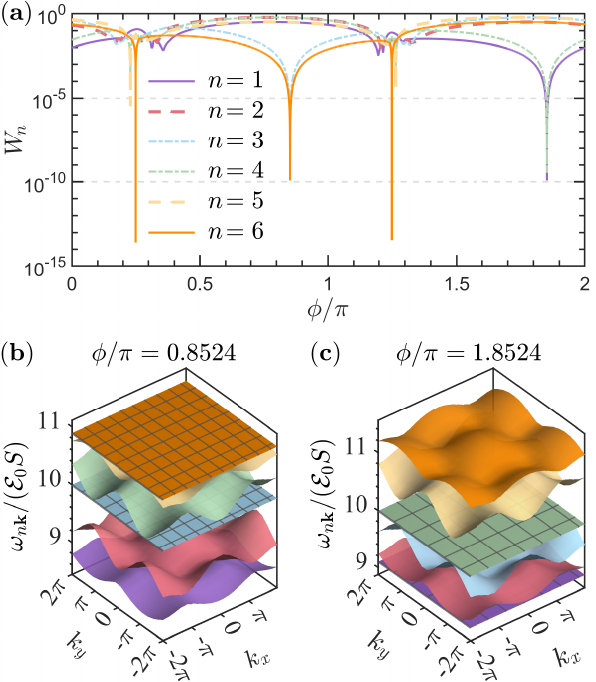}\\
\caption{(a) The bandwidth $W_n$ ($n = 1, 2, \cdots, 6$) in the Kitaev-$\Gamma$ model at a magnetic field strength of   $h/(\mathcal{E}_0 S) = 10$. In the plot, the increment of exchange coupling angle is $\delta\phi/\pi$ = 0.0001. The pronounced drops are located at $\phi/\pi \simeq 1/4$ and $5/4$, $\phi/\pi \approx$ 0.8524 and 1.8524, and $\phi/\pi \approx$ 0.2296 and 1.2643. 
  (b) When $\phi/\pi \approx 0.8524$, the flat bands occur at the third and the sixth energy bands. (c) When $\phi/\pi \approx 1.8524$, the flat bands occur at the first and the fourth energy bands.
  }\label{FIG-Flatband}
\end{figure}

By tuning $\phi$ from 0 to $2\pi$, we have calculated the bandwidths $W_n$ of the six energy bands at fixed $h/(\mathcal{E}_0S) = 10$, see Fig.~\ref{FIG-Flatband}(a).
The most striking observation is that, in addition to the topological phase-transition points occurring at $\phi/\pi \simeq 1/4$ and $5/4$, the bandwidths of certain energy bands become vanishingly narrow at specific parameter values.
These include bands 3 and 6 at $\phi/\pi \approx 0.8524$, and bands 1 and 4 at $\phi/\pi \approx 1.8524$.
The bandwidths ($\sim10^{-10}$) therein are ten orders of magnitude smaller than the characteristic energy scale $\mathcal{E}_0 = 1$.
As can be seen in Fig.~\ref{FIG-Flatband}(b) at $\phi/\pi \approx 0.8524$,
the third and sixth bands are discerned as the flat bands.
Both flat bands acquire finite Chern numbers of $-1$, endowing them as topological flat bands. It is also expected that the flat bands have an extremely high density of states that is akin to the van Hove singularity.
However, the flat bands remain poorly isolated due to the extremely small (albeit finite) band gaps.
Similarly, from Fig.~\ref{FIG-Flatband}(c) it is observed that the numerical orders of the flat bands at $\phi/\pi \approx 1.8524$ are the first and the fourth, and their associated Chern numbers are $+1$.
Additionally, for the exchange coupling angle $\phi/\pi \approx$ 0.2296 and 1.2643, we also identified flat-band structures in the fourth and fifth bands.
Since both exchange coupling angles belong to Phase II, the Chern numbers of the fourth and fifth flat bands are opposite and are $+1$ and $-1$, respectively.
For details of the density of states and the extra energy-momentum dispersion, see Figs. S3 and S4 in the SM \cite{SuppMat}.

\begin{figure}[!ht]
\centering
\includegraphics[width=0.95\columnwidth, clip]{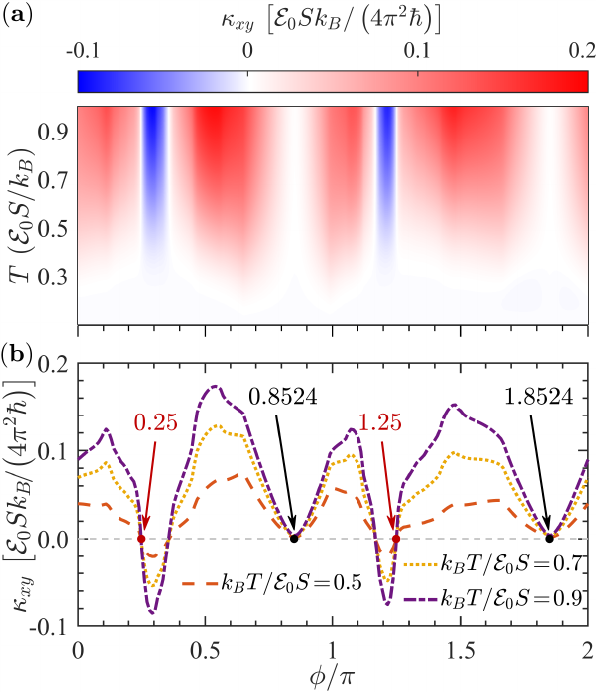}\\
\caption{ The thermal Hall conductivity $\kappa_{xy}$ in the 
  Kitaev-$\Gamma$ model at a shifted magnetic field strength of $\delta h/(\mathcal{E}_0 S) = 1$ when compared to the magnetic phase boundary.
  (a) Contour plot of $\kappa_{xy}$ as a function of exchange coupling angle $\phi$ and temperature $T$.
  (b) $\kappa_{xy}$ as a function of $\phi$ at different $T$. The red and black arrows indicate the phase-transition points and flat-band points, respectively. Here we set $k_B = \hbar = 1$.
  }\label{FIG-THE}
\end{figure}

%%============================================================================
\subsection{Nodes in thermal Hall conductivity}
%%============================================================================

Until now, thermal Hall conductivity has emerged as a crucial transport diagnostic tool for probing emergent phenomena in Kitaev materials (for a latest review, see Ref.~\cite{Zhang2024PP}).
Although a topologically trivial phase may theoretically yield a nonzero value, thermal Hall conductivity is predominantly ascribed to topological phases with finite Chern numbers.
In addition, thermal Hall conductivity can also serve as a sensitive probe to pinpoint topological phase transitions.
For this purpose, we again focus on the Kitaev-$\Gamma$ model parameterized by a exchange coupling angle $\phi$, and investigate the thermal Hall conductivity $\kappa_{xy}$ along a special path in which the magnetic field is $\phi$-dependent and is larger than the corresponding critical magnetic field defined in Eq.~\eqref{EQ:PhaseBound} and Table~\ref{Tab-FitingFormula} by 1.
This strategy ensures that $\kappa_{xy}$ remains stable throughout the full parameter range, without experiencing significant fluctuations.

Figure~\ref{FIG-THE}(a) shows the contour plot of thermal Hall conductivity $\kappa_{xy}$ as a function of temperature $T$ in the range of $\phi \in [0, 2\pi)$.
The $\kappa_{xy}$ is basically positive and is intervened by two negative regions, 
giving rise to four sign-changing lines (white segment) which are almost temperature independent when $T \gtrsim 0.3$.
In addition, there are two extra diffusive lines at which $\kappa_{xy}$ is vanishingly small.
These distinct lines determine the nodes in the thermal Hall conductivity.

To elucidate the nature of these nodes, Fig.~\ref{FIG-THE}(b) presents behaviors of $\kappa_{xy}$ at three distinct temperatures.
It is evident that as the temperature $T$ increases, the magnitude of $\kappa_{xy}$ has a general incline, except for six parameter points where $\kappa_{xy}$ remains near zero.
When $\phi/\pi = 1/4$ and $5/4$, there are bands touching in the dispersion relation (cf. Fig. S5 in the SM \cite{SuppMat}), rendering the Chern numbers ill-defined.
Interestingly, the band-touching phenomenon is robust against the magnetic field.
Recalling that the two exchange coupling angles represent hidden SU(2) Heisenberg points in the honeycomb-lattice Kitaev-$\Gamma$ model \cite{Giniyat2015PRB}, we speculate that the same ansatz may still occur on the star lattice.
Therefore, in these cases the nodes are indicators of topological phase transitions due to intrinsic band touching.
Furthermore, the nodes at $\phi/\pi \approx 0.3558$ and 1.1712 are more involved. For the parameters, the underlying Chern number tuples is $(1,1,-3,3,-1,-1)$. The major contributions of $\kappa_{xy}$ come from higher Chern numbers $C = \pm 3$. However, the corresponding $\kappa_{xy}$ cancel out considerably because of opposing Chern numbers.
This further leads to the general cancelation of all branches.
However, as we change the increment of magnetic field from the magnetic phase boundary, 
it is found that the positions of the nodes can vary and they approach their neighboring $\phi/\pi = 1/4$ or $5/4$.
Intriguingly, the pattern of the underlying Chern number tuple persists regardless of the magnetic field (for details, see Fig. S6 in the SM \cite{SuppMat}). 
Considering that these intermediate topological phases gradually vanish as the magnetic field increases, the nodes are interpreted as the onset of topological fragments exemplified by the Chern number tuple $(1,1,-3,3,-1,-1)$.
Therefore, their occurrence leads to sign reversals of the thermal Hall conductivity near the topological phase transitions.

Moreover, the locations of the flat-band points, $\phi/\pi \approx 0.8524$ and 1.8524, coincide precisely with the thermal Hall conductivity approaching zero, and no sign change in $\kappa_{xy}$ is observed in the vicinity of these points due to the lack of a topological phase transition.
The reason for the vanishingly small value of $\kappa_{xy}$ is analyzed as follows.
Taking the parameter $\phi/\pi \approx 0.8524$ whose Chern number tuple is equal to $(-1,1,-1,0,2,-1)$ as an example, 
the Berry curvatures of the two lowest bands have different signs and are strongly localized around the $\textbf{K}$ and $\textbf{K}'$ points in the Brillouin zone. This causes the contributions from the two lowest bands to $\kappa_{xy}$ to nearly cancel out.
On the other hand, $\kappa_{xy}$ of the energy band with the highest Chern number ($+2$), together with the correlation of the energy band with zero Chern number, is approximately counterbalanced by its neighboring flat bands with $C = -1$.
Such an offset may stem from the distinctive role of the flat bands, which exhibit vanishing group velocities, along with their unique population and coupling with other dispersive bands.
Together, these pivots lead to almost zero thermal Hall conductivity.
As the exchange coupling angle deviates slightly from the flat-band points, $\kappa_{xy}$ increases parabolically (Numerical details can be found in Figs. S7 and S8 in the SM \cite{SuppMat}).
However, the fundamental mechanism underlying the reduction of thermal Hall conductivity in flat-band regions remains elusive, warranting further theoretical study.

%%%%%%%%%%%%%%%%%%%%%%%%%%%%%%%%%%%%%%%%%%%%%%%%%%%%%%%%%%%%%%%%%%%%%%%%%%%%%%
\section{Conclusion}\label{SEC:CONC}
%%%%%%%%%%%%%%%%%%%%%%%%%%%%%%%%%%%%%%%%%%%%%%%%%%%%%%%%%%%%%%%%%%%%%%%%%%%%%%

In summary, we have studied the unusual band topology of the field-polarized phase in the $JK\Gamma\Gamma'$ model on the star lattice using linear spin-wave theory.
Upon conducting a meticulous exploration within the large parameter space, we have successfully discerned the presence of two distinct topological phases with high Chern number $C = \pm2$. Additionally, many flimsy high-Chern-number phases are achieved on the boundary of the two overwhelmingly topological phases.
In the paradigmatic Kitaev-$\Gamma$ model, these intermediate topological phases vanish at high enough magnetic field, and their direct topological phase transitions occur at $\phi \simeq \pi/4$ and $5\pi/4$, respectively. 
In addition, we strive to identify a pair of topological flat bands within each of the two distinct phases.
Interestingly, the thermal Hall conductivity exhibits nodes at both the topological phase-transition points and the flat-band points. However, it only experiences sign changes at the former rather than the latter.

For comparison, while the high-field topological phase characterized by a distinct Chern number tuple $(C_1, C_2) = (+1, -1)$ has been observed in honeycomb-lattice Kitaev magnets, our findings here exhibit significant differences from these previously reported counterparts. Notably, our results demonstrate three key distinguishing features: the emergence of higher Chern numbers, the formation of topological flat bands, and the occurrence of multiple topological phase transitions. These observations collectively highlight the crucial interplay between exchange couplings and geometric frustration in governing the topological properties of the system.

In short, our findings on the star lattice provide insights for high Chern numbers and topological flat bands in Kitaev materials, and they also deepen the understanding of oscillations in thermal Hall conductivity.
Nevertheless, further investigation is warranted to search for stable topological phases with higher Chern numbers ($|C| > 2$) and explore the impacts of magnon-magnon interactions on the topology of flat bands \cite{McClarty2018PRB,Mook2021PRX,Maksimov2022PRB,Chen2023PRB,Wang2025arXiv}.
It is also imperative to map out the zero-field magnetic phase diagrams for the benefit of uncovering possible hidden SU(2) points akin to those in the honeycomb lattice \cite{Giniyat2015PRB}.

\begin{acknowledgements}
We thank K. Chen, X. Li, and C. Wang for helpful discussions.
This work is supported by the National Natural Science Foundation of China (Grants No. 12304176 and No. 12274183)
and the Natural Science Foundation of Jiangsu Province (Grant No. BK20220876).
The computations are partially supported by High Performance Computing Platform of Nanjing University of Aeronautics and Astronautics.
\end{acknowledgements}

%%%%%%%%%%%%%%%%%%%%%%%%%%%%%%%%%%%%%%%%%%%%%%%%%%%%%%%%%%%%%%%%%%%%%%%%%%%%%%%%%%%%%%%%%%%%%%%%%%%%%%%%%

%
%%%%%%%%%%%%%%%%%%%%%%%%%%%%%%%%%%%%%%%%%%%%%%%%%%%%%%%%%%%%%%%%%%%%%%%%%%

%%%%%%%%%%%%%%%%%%%%%%%%%%%%%%%%%%%%%%%%%%%%%%%%%%%%%%%%%%%%%%%%%%%%%%%%%%%%%%%%%%%%%%%%%%%%%%%%%%%%%%%%%%%%%%%%%%%%%%%%%%%%%%%%%%%%%%%%%%%%%%%%

%%%%%%%%%% Merge with supplemental materials %%%%%%%%%%
%% ref: https://arxiv.org/abs/1709.10096

%%%%%%%%%%%%%%%%%%%%%%%%%%%%%%%%%%%%%%%%%%%%%%%%%%%%%%%%%%%%%%%%%%%%%%%%%%%%%%%%%%%%%%%%%%%%%%%%%%%%%%%%%%%%%%%%%%%%%%%%%%%%%%%%%%%%%%%%%%%%%%%%

\clearpage

\onecolumngrid

%%%%%%%%%% Merge with supplemental materials %%%%%%%%%%
%%%%%%%%%% Prefix a "S" to all equations, figures, tables and reset the counter %%%%%%%%%%
\newpage

\newcounter{sectionSM}
\newcounter{equationSM}
\newcounter{figureSM}
\newcounter{tableSM}
\stepcounter{equationSM}
\setcounter{section}{0}
\setcounter{equation}{0}
\setcounter{figure}{0}
\setcounter{table}{0}
\setcounter{page}{1}
\makeatletter
\renewcommand{\thesection}{\textsc{S}\arabic{section}}
\renewcommand{\theequation}{\textsc{S}\arabic{equation}}
\renewcommand{\thefigure}{\textsc{S}\arabic{figure}}
\renewcommand{\thetable}{\textsc{S}\arabic{table}}

% \onecolumngrid

%%%%%%%%%%%%%%%%%%%%%%%%%%%%%
\begin{center}
{\large{\bf Supplemental Material for\\
``High Chern numbers and topological flat bands\\ in high-field polarized Kitaev magnets on the star lattice''}}
\end{center}
\begin{center}
Zixuan Zou$^{1}$ and Qiang Luo$^{1,\;2}$ \\
\quad\\
$^1$\textit{College of Physics, Nanjing University of Aeronautics and Astronautics, Nanjing, 211106, China}\\
$^2$\textit{Key Laboratory of Aerospace Information Materials and Physics (NUAA), MIIT, Nanjing, 211106, China}\\
(Dated: June 4th, 2025)
\quad\\
\end{center}

%\begin{widetext}
%\end{widetext}

% \twocolumngrid
\onecolumngrid

%%%%%%%%%%%%%%%%%%%%%%%%%%%%%

% In this supplemental material~(SM), we present some additional figures for a better understanding of the results in the main text.
% Figure S1: topological phase diagrams of the $JK\Gamma$ model ($J < 0$) and $K\Gamma\Gamma'$ ($K < 0$) model with $h/(\mathcal{E}_0S) = 200$, 2) topological phase diagrams of the Kitaev-$\Gamma$ model with $h/(\mathcal{E}_0S)$ = 150, 100, and 50, 3) density of states of the Kitaev-$\Gamma$ model at $\phi \approx 0.8444\pi$ and $1.8444\pi$, 4) flat bands of the Kitaev-$\Gamma$ model at $\phi \approx 1.2667\pi$, 5) dispersion relation of the Kitaev-$\Gamma$ model at $\phi = \pi/4$ and $5\pi/4$, and 6) thermal Hall conductivity near the variable nodes with $|C| = 3$.

\vspace{-0.00cm}
\section{Supplementary Figure(s) for Sec. III.A of the Main Text}

\begin{figure}[!ht]
\centering
\includegraphics[width=0.82\columnwidth, clip]{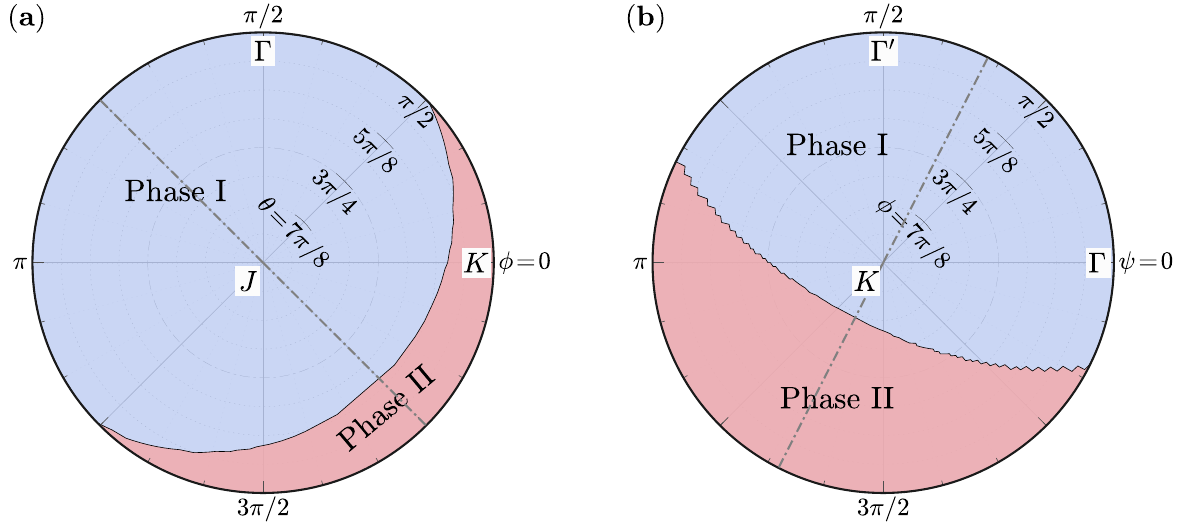}\\
\caption{(a) and (b) show the topological phase diagrams of the $JK\Gamma$ model (with $\psi = 0$) and $K\Gamma\Gamma'$ model (with $\theta = \pi/2$). Different from the Fig. 2 shown in the main text, range of $\theta$ are $\phi$ is $[\pi/2, \pi]$, such that $J \leq 0$ and $K \leq 0$, respectively.
  }
\end{figure}

\vspace{-0.00cm}
\section{Supplementary Figure(s) for Sec. III.B of the Main Text}

\begin{figure}[!ht]
\centering
\includegraphics[width=0.80\columnwidth, clip]{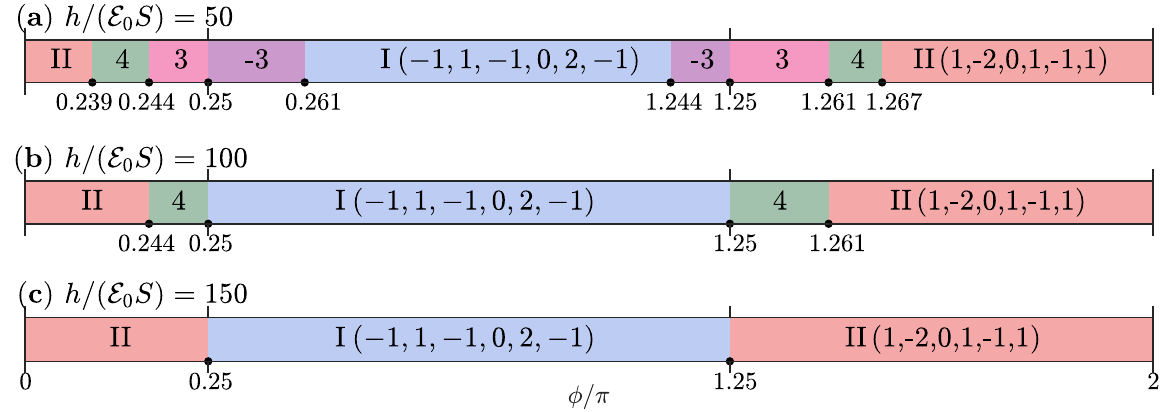}\\
\caption{The topological phase diagrams of the Kitaev-$\Gamma$ model at different magnetic fields, which include (a) $h/(\mathcal{E}_0S) = 50$, (b) $h/(\mathcal{E}_0S) = 100$, and (c) $h/(\mathcal{E}_0S) = 150$, respectively. The phases are marked by their largest Chern number in the sense of absolute value. In addition to topological phases I and II, the Chern number tuples for the other phases are $-3$ for $(1,1,-3,3,-1,-1)$, $3$ for $(-1,3,-3,0,0,1)$, and $4$ for $(-2,4,-3,0,0,1)$.
  }
\end{figure}

\begin{figure}[!ht]
\centering
\includegraphics[width=0.55\columnwidth, clip]{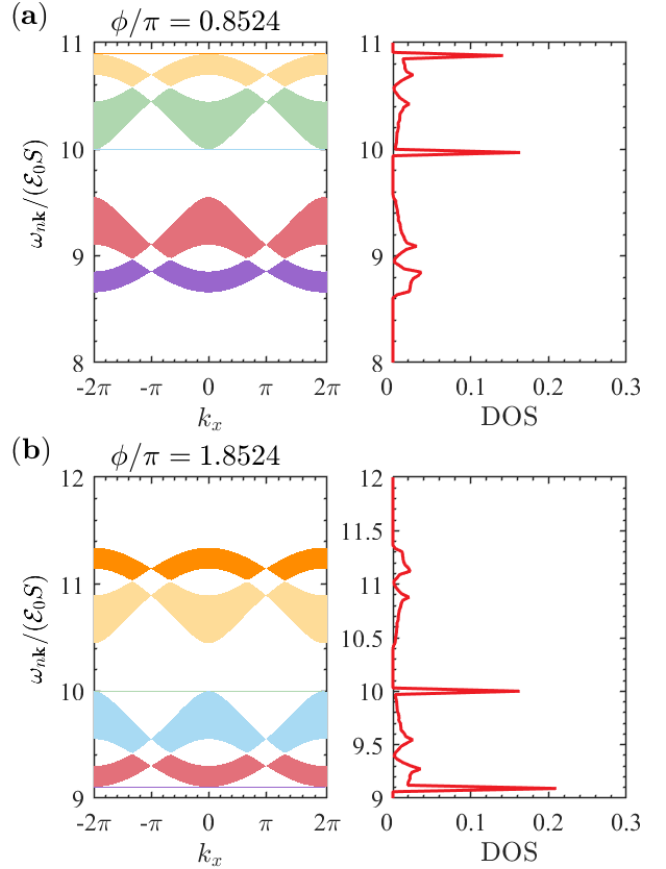}\\
\caption{(a) The front view of the energy-momentum dispersion and the corresponding density of states for the the Kitaev-$\Gamma$ model at $\phi/ \pi \approx 0.8524$. (b) The similar plot for (a) but with $\phi/ \pi \approx 1.8524\pi$.
  }
\end{figure}

\begin{figure}[!ht]
\centering
\includegraphics[width=0.55\columnwidth, clip]{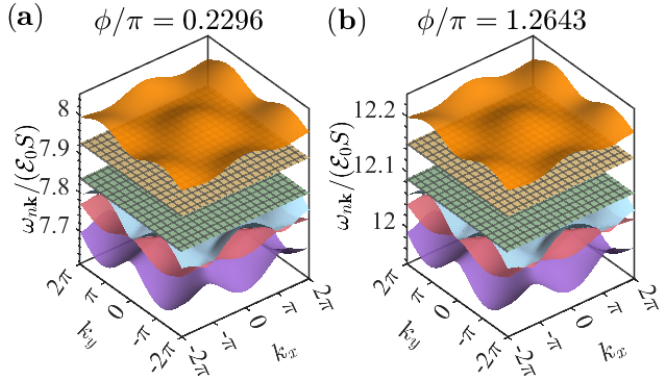}\\
\caption{The topological flat bands in the Kitaev-$\Gamma$ model at a magnetic field strength of $h/(\mathcal{E}_0 S) = 10$. In panel (a) and (b), $\phi/\pi \approx$ 0.2296 and 1.2643, respectively. The flat bands occur at the fourth ($C_4 = +1$) and the fifth ($C_5 = -1$) energy bands.}
\end{figure}

\clearpage
\section{Supplementary Figure(s) for Sec. III.C of the Main Text}

\begin{figure}[!ht]
\centering
\includegraphics[width=0.55\columnwidth, clip]{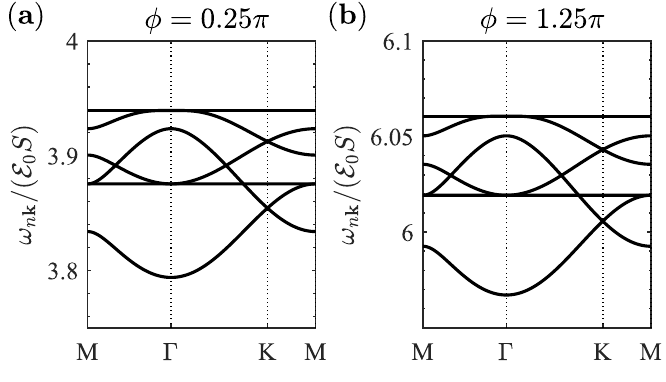}\\
\caption{Magnon band structures for the Kitaev-$\Gamma$ model at a magnetic field strength $h/(\mathcal{E}_0 S) = 10$. (a) and (b) are for $\phi/\pi = 1/4$ and $\phi/\pi = 5/4$, respectively.
  }
\end{figure}

\begin{figure}[!ht]
\centering
\includegraphics[width=0.95\columnwidth, clip]{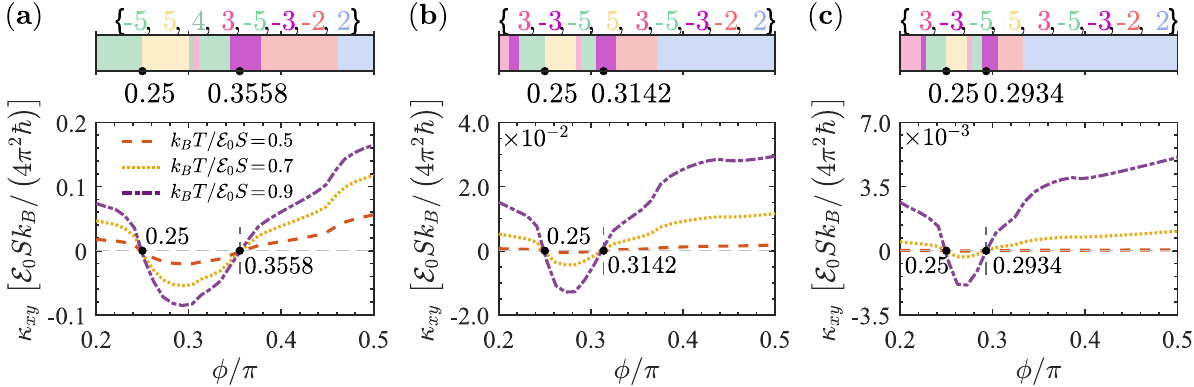}\\
\caption{(a) The topological phase diagram and the thermal Hall conductivity $\kappa_{xy}$ in the 
  Kitaev-$\Gamma$ model at a shifted magnetic field strength of $\delta h/(\mathcal{E}_0 S) = 1$ when compared to the magnetic phase boundary shown in Fig.~4 in the main text.
  It is observed that the parameter $\phi/\pi$ at which $\kappa_{xy} = 0$ is 0.3552, and it belongs to the topological phase whose Chern number tuple is $(1,1,-3,3,-1,-1)$.
  (b) The shift magnetic field is 3, and the parameter $\phi/\pi$ at which $\kappa_{xy} = 0$ is 0.3138. (c) The shift magnetic field is 5, and the parameter $\phi/\pi$ at which $\kappa_{xy} = 0$ is 0.2931. 
  }
\end{figure}

\begin{figure}[!ht]
\centering
\includegraphics[width=0.90\columnwidth, clip]{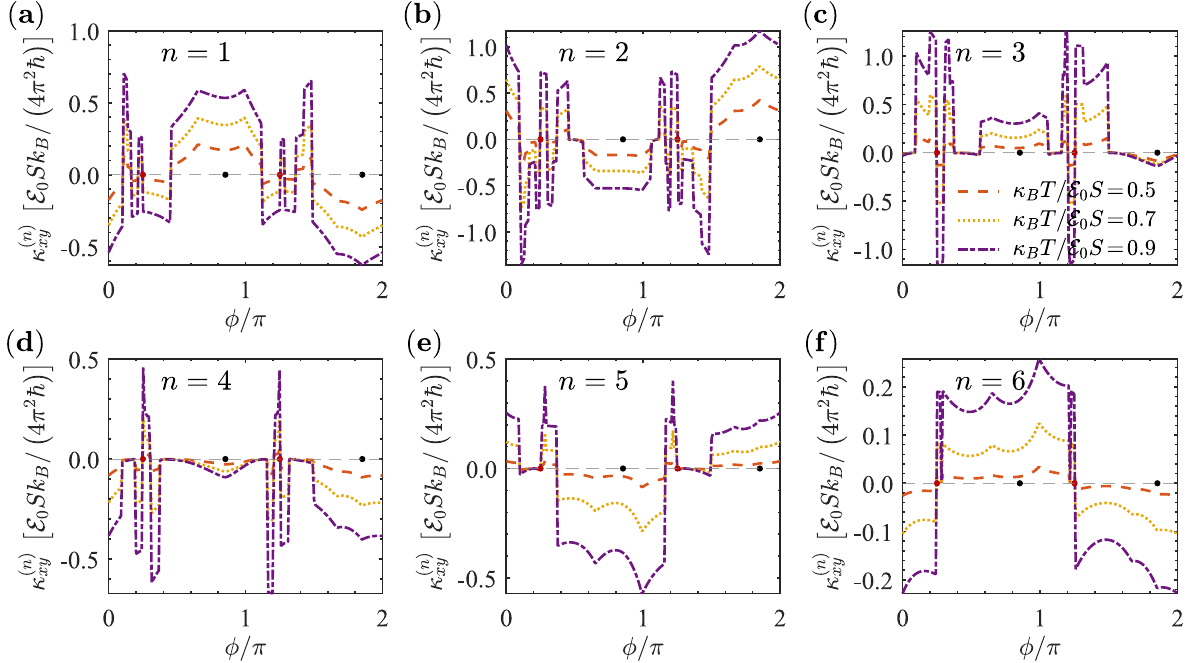}\\
\caption{(a)-(f) shows the thermal Hall conductivity $\kappa_{xy}^{(n)}$ of the $n$-th energy band ($n = 1, 2, \cdots, 6$) in the Kitaev-$\Gamma$ model at a shifted magnetic field strength of $\delta h/(\mathcal{E}_0 S) = 1$ when compared to the magnetic phase boundary. The entire thermal Hall conductivity $\kappa_{xy} = \sum_{n=1}^6 \kappa_{xy}^{(n)}$ is shown in Fig.~\textcolor{red}{7}(b) in the main text. The red and black dots indicate the phase-transition points ($\phi/\pi \simeq 1/4$ and $5/4$) and flat-band points ($\phi/\pi \approx$ 0.8524 and 1.8524), respectively.}\label{FIGSM-KxyBand}
\end{figure}

\begin{figure}[!ht]
\centering
\includegraphics[width=0.90\columnwidth, clip]{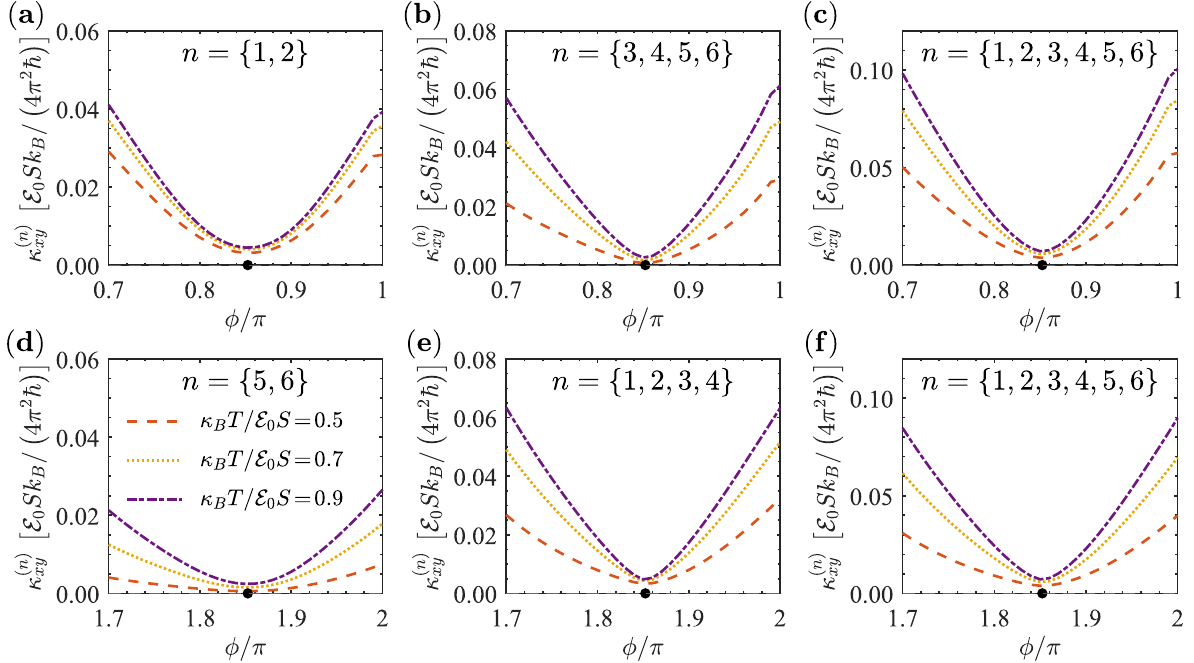}\\
\caption{The proper combination of thermal Hall conductivity $\kappa_{xy}^{(n)}$ in the base of Fig.~\ref{FIGSM-KxyBand}. 
(a)-(c): For the flat-band point $\phi/\pi \approx 0.8524$, the Chern number tuple is $(-1, 1, -1, 0, 2, -1)$ and the third and sixth energy bands are flat. Thus, panel (a) shows the sum of $\kappa_{xy}^{(n)}$ for the lowest two bands with $n = \{1, 2\}$, while panel (b) shows the sum for the remaining four bands with $n = \{3, 4, 5, 6\}$. Panel (c) shows the entire $\kappa_{xy}$, which increases parabolically when away from the flat-band point. 
(d)-(f): Similar to (a)-(c) but for flat-band point $\phi/\pi \approx 1.8524$, at which the Chern number tuple is $(1, -2, 0, 1, -1, 1)$ and the first and fourth energy bands are flat.
  }
\end{figure}

% \end{CJK*}

\end{document}